\documentclass[11pt]{article}

\usepackage{hyperref}
\hypersetup{
    colorlinks=true,
    linkcolor=blue
    }

\usepackage{latexsym,color,graphicx,amsmath,amssymb}
\usepackage[margin=1.25in]{geometry}

\usepackage{mathpazo}
\usepackage{calrsfs}

\usepackage{cite}
 \usepackage[amsmath,thmmarks]{ntheorem}%
 \theoremseparator{.}

\def\reals{{\mathbb R}}
\def\eps{{\varepsilon}}

\def\A{\mathcal{A}}
\def\B{\mathcal{B}}
\def\C{\mathcal{C}}

\def\F{\mathcal{F}}
\def\G{\mathcal{G}}

\def\XX{\mathsf{X}}
\def\xx{{\bf x}}
\def\yy{{\bf y}}
\def\sd{{s}}

\def\trap{\psi}
\def\trapS{\Psi}
\def\biclique{\B}
\def\rangeS{\Sigma}
\def\srange{\sigma}
\def\store{\omega}
\def\slab{W}

\def\ceil#1{\lceil #1\rceil}
\def\cell{\tau}
\def\cuthierarchy{\boldsymbol{\Xi}}
\def\parent{\mathsf{p}}

\def\etal{\textsl{et~al.}}

\newtheorem{theorem}{Theorem}[section]
\newtheorem{lemma}[theorem]{Lemma}
\newtheorem{corollary}[theorem]{Corollary}
 
\newcommand{\myqedsymbol}{\rule{2mm}{2mm}}

\theoremheaderfont{\em}%
\theorembodyfont{\upshape}%
\theoremstyle{nonumberplain}%
\theoremseparator{}%
\theoremsymbol{\myqedsymbol}%
\newtheorem{proof}{Proof:}%

\newtheorem{proof:sketch}{Proof (sketch):}%

\title{Semi-Algebraic Off-line Range Searching and Biclique Partitions in the Plane
\thanks{%
  A preliminary version of this paper appeared in the \emph{Proc. 40th International Symposium on Computational Geometry, 2024, 4:1--4:15.}\newline
  Work by Pankaj Agarwal has been partially supported by NSF grants CCF-20-07556, CCF-22-23870, and IIS-24-02823, and by the Binational Science Foundation Grant 2022131.
  Work by Esther Ezra has been partially supported by Israel Science Foundation Grant 800/22,
  and Binational Science Foundation Grant 2022131. 
  Work by Micha Sharir has been partially supported by Israel Science Foundation Grant 495/23.}
  }

\author{%
	Pankaj K. Agarwal\thanks{
    Department of Computer Science, Duke University, Durham, NC 27708, USA;
    {\sf pankaj@cs.duke.edu,
    https://orcid.org/0000-0002-9439-181X}
    }
    \and
    Esther Ezra\thanks{
    Department of Computer Science, Bar Ilan University, Israel
     {\sf ezraest@cs.biu.ac.il,
      https://orcid.org/0000-0001-8133-1335}
    }
  \and
    Micha Sharir\thanks{
    School of Computer Science, Tel Aviv University, Tel Aviv, Israel;
    {\sf michas@tauex.tau.ac.il,
      http://orcid.org/0000-0002-2541-3763}}
}

\begin{document}

\maketitle

\begin{abstract}
  Let $P$ be a set of $m$ points in $\reals^2$, let $\rangeS$ be a set of $n$
  semi-algebraic sets of constant complexity in $\reals^2$, let $(S,+)$ be a semigroup, and let 
  $w: P \rightarrow S$ be a weight function on the points of $P$. We describe a randomized algorithm for 
  computing $w(P\cap\srange) = \sum_{p\in P\cap\sigma} w(p)$ for every $\srange\in\rangeS$ in overall expected time
  $O^*\bigl( m^{\frac{2s}{5s-4}}n^{\frac{5s-6}{5s-4}} + m^{2/3}n^{2/3} + m + n \bigr)$,
  where $s>0$ is the number of degrees of freedom of the regions of $\Sigma$, and where the 
  $O^*(\cdot)$ notation hides subpolynomial factors. For $s\ge 3$, surprisingly, this bound is 
  smaller than the best-known bound for answering $m$ such queries in an on-line manner; 
  the latter takes $O^*(m^{\frac{s}{2s-1}}n^{\frac{2s-2}{2s-1}}+m+n)$ time. 
  
  Let $\Phi: \rangeS \times P \rightarrow \{0,1\}$ be the Boolean predicate (of constant complexity) such 
  that $\Phi(\srange,p) = 1$ if $p\in\srange$ and $0$ otherwise, and let $\rangeS\mathop{\Phi} P = \{ (\srange,p) \in \rangeS\times P 
  \mid \Phi(\srange,p)=1\}$.
  Our algorithm actually computes a partition $\biclique_\Phi$ of $\rangeS\mathop{\Phi} P$ into 
  (edge-disjoint) bipartite cliques (bicliques) of size (i.e., sum of the sizes of the vertex sets of its bicliques) 
  $O^*\bigl( m^{\frac{2s}{5s-4}}n^{\frac{5s-6}{5s-4}} + m^{2/3}n^{2/3} + m + n \bigr)$.
  It is straightforward to compute $w(P\cap\srange)$ for all $\srange\in \rangeS$ from $\biclique_\Phi$, 
  in either off-line or on-line manner (so the only off-line component of our algorithm is the construction of the biclique partition).
  Similarly, if $\eta: \Sigma \rightarrow S$ is a weight function on the regions of $\Sigma$, 
  $\sum_{\srange\in \rangeS: p \in \srange} \eta(\srange)$, for every point $p\in P$, can be computed from 
  $\biclique_\Phi$ in a straightforward manner, in the same asymptotic time bound, again either off-line or on-line. A recent work of Chan et al.~\cite{CCZ24} solves the on-line version
  of this dual \emph{point enclosure} problem within the same performance bound as our off-line solution.
  We also mention a few other applications of computing $\biclique_\Phi$.
\end{abstract}

\section{Introduction}

A typical range-searching problem asks to preprocess a set $P$ of $m$ points in $\reals^d$ into a data 
structure so that, for a query region $\srange$, some aggregate statistics on $\srange \cap P$ can be computed quickly, e.g.,  testing whether $\srange\cap P = \emptyset$, computing $|\srange\cap P|$, or computing a weighted sum of $\srange\cap P$ (given a weight function on $P$ taking values in some semigroup).
A central problem in computational geometry, 
range searching has been extensively studied over the last five decades, and sharp bounds are known 
for many instances; see~\cite{AC21,AC22,Ag,AE99,Mat94} and references therein. For instance, a simplex range 
query (where the query region is a simplex) can be answered in $O^*(m/\store^{1/d})$ time using $O^*(\store)$ space 
and preprocessing for any $\store\in[m,m^d]$, and (almost) matching lower bounds are known.\footnote{%
  Throughout this paper, the $O^*(\cdot)$ notation hides subpolynomial factors, typically of the form $m^\eps$, 
  and its associated $\eps$-dependent constant of proportionality, for any $\eps>0$.}
In particular, with a suitable choice of $\store$, the total time spent, including the preprocessing cost, in answering a set $\rangeS$ of 
$n$ simplex range queries is
$O^*((mn)^{1-\tfrac{1}{d+1}}+m+n)$. The known lower bounds imply that this bound is tight within a $\log^{O(1)}m$ factor. 
However, such sharp upper and lower bounds are not known for more general classes of range queries. For instance, 
the best known data structures answer a disk range query (for points in the plane and disks as queries) in
$O^*(m^{3/4}/\store^{1/4})$ time
using $O(\store)$ space and $O^*(\store)$ preprocessing, for any 
$\store\in [m,m^3]$, and thus the total cost of answering $n$ disk range queries is $O^*(m^{3/5}n^{4/5}+m+n)$,
while the best known 
lower bound is $\Omega(m^{2/3}n^{2/3})$.
(Slightly better lower bounds are known for annulus range queries~\cite{AC21}.)
A similar gap holds (see below for the exact bounds) for  the more general class of semi-algebraic range 
queries.\footnote{%
  Roughly speaking, a semi-algebraic set in $\reals^d$ is the set of points 
  in $\reals^d$ satisfying a Boolean predicate over a set of real polynomial inequalities; 
  the complexity of the predicate and of the set is defined in terms of the number of 
  polynomials involved and their maximum degree; see \cite{BPR} for details.}
A natural and fundamental open question is whether this gap can be narrowed. There is some evidence that the 
current upper bounds are not optimal.

Given a set $P$ of $m$ points and a set $\Gamma$ of $n$ surfaces in $\reals^d$, the 
\emph{incidence problem} on $P$ and $\Gamma$ asks for obtaining a sharp bound on the maximum 
number of \emph{incidences} between these sets, i.e., pairs $(p,\gamma) \in P\times \Gamma$ such 
that $p\in\gamma$. Originally posed for bounding the number of incidences between points and lines in the plane
\cite{ST83}, by now there is vast literature on this topic; 
  see~\cite{ANPPSS,PS04,Sh22,SZ,Sze,ST83} for a sample of references.
  There is a deep connection between range searching and the incidence problem.
  For example, many of the techniques developed for bounding incidences 
  (e.g., geometric cuttings and polynomial partitioning techniques)
  have led to fast data structures for range searching, and vice versa.
  Similarly, many of the lower-bound constructions for range searching exploit the 
  incidence structure between points and curves/surfaces~\cite{AC21}.
  As such, there is a general belief that the two problems are closely related, and that the upper bound on 
  the running time of (at least off-line) range queries should be almost the same as the upper bound on the number of incidences 
  between points and the corresponding curves/surfaces that bound the query regions. This certainly holds for simplex (triangle) range searching and for 
  incidences between points and lines in $\reals^2$, and, with some constraints, for points and halfspaces 
  (for range searching) or hyperplanes (for incidences) 
  in higher dimensions; see, e.g.,~\cite{ApS07,BK}. This also used to be the case for disk range searching and 
  point-circle incidence problem --- the best known upper bound on incidences between 
  $m$ points and $n$ circles used to be $O(m^{3/5}n^{4/5}+m+n)$ (see, e.g., Pach and Sharir~\cite{PS}). However,
  Aronov and Sharir~\cite{ArS:circ}, and later Agarwal et al.~\cite{ANPPSS}, obtained an improved bound of 
  $O^*(m^{2/3}n^{2/3}+m^{6/11}n^{9/11}+m+n)$ for point-circle incidences
  (see also~\cite{AAS03,SZ} for related results), and later Agarwal and Sharir~\cite{AS05} presented 
  an algorithm for computing these incidences in the same time bound (up to an additional $\log n$ 
  factor in the $O^*(\cdot)$ notation).
  More recently, Sharir and Zahl~\cite{SZ} obtained a bound of $O^*(m^{\frac{2s}{5s-4}}n^{\frac{5s-6}{5s-4}} + m^{2/3}n^{2/3} + m + n)$ 
  on the number of incidences between $m$ points and $n$ semi-algebraic curves of constant complexity, where $s$ is 
  the number of degrees of freedom of the curves (the number of real parameters needed to specify a curve).
  If we believe the above conjecture, as we tend to, a natural question is whether one can obtain
  algorithms for disk range searching, and more broadly for semi-algebraic range searching, that have these 
  running times, up to possible $O^*(\cdot)$ factors, at least in the off-line setting.

In this paper we answer this question in the affirmative for $d=2$, by presenting
an algorithm for the off-line semi-algebraic range-searching  problem in $\reals^2$, with (randomized
expected) running time that almost matches (again, up to $O^*(\cdot)$ factors) the aforementioned 
incidence bounds. Our algorithm also works for off-line point-enclosure queries (see below) amid semi-algebraic sets 
in $\reals^2$ within the same time bound. As already mentioned in the abstract and will be discussed later,
a recent result of Chan et al.~\cite{CCZ24} shows that for point enclosure queries (but not for range searching queries), answering $n$ such queries in
an on-line context can be performed within the same bound as in the off-line setting discussed in this paper.

\paragraph{Problem statement.}
Let $P$ be a set of $m$ points and let $\rangeS$ be a set of $n$ semi-algebraic sets
of constant complexity in $\reals^2$. Let $s$ denote the \emph{parametric dimension} (also known as the number of
\emph{degrees of freedom}) of the regions 
in $\rangeS$, for some constant $s>0$, meaning that each region can be specified by at most $s$ real 
parameters. Let $(S,+)$ be a semigroup, and let $w: P \rightarrow S$ be a weight function. For a subset 
$R\subseteq P$, let $w(R) = \sum_{p\in R} w(p)$. Our goal is to compute $w(P\cap\srange)$, for every $\srange\in\rangeS$. 
As already mentioned, this semigroup model encapsulates many popular variants of range searching~\cite{Ag}. Alternatively, 
we may assign a weight function $\eta: \rangeS \rightarrow S$ and compute, for every point $p\in P$, 
the weight $\sum_{\srange\in\rangeS: p\in \srange} \eta(\srange)$. This dual setup is referred to as
\emph{point enclosure} searching.

To solve the above problems, and some of their variants, we formulate a more general problem:
Let $\Phi: \rangeS \times P \rightarrow \{0,1\}$ be the predicate such that,
for $\srange\in\rangeS$ and $p \in P$ (more generally, for any range $\sigma$ of the kind considered in $\Sigma$ and for any point $p$), $\Phi(\srange, p) =1$ iff $p\in\srange$.
$\Phi$ is a semi-algebraic predicate, defined as a Boolean combination of a constant number of real polynomial inequalities , and is of constant complexity, meaning that it involves a constant number of polynomials of constant maximum degree.
Let $\rangeS \mathop{\Phi} P = \{ (\srange, p) \in \rangeS\times P \mid \Phi(\srange,p)=1\}$. A popular 
method of representing
$\rangeS \mathop{\Phi} P$ compactly is to use a \emph{biclique partition} 
$\biclique_\Phi := \biclique_\Phi(\rangeS,P) = \{ (\rangeS_1, P_1), \ldots, (\rangeS_u, P_u)\}$, 
where $\Phi(\srange, p) = 1$ for all indices $i$ and for all pairs 
$(\srange, p) \in \rangeS_i \times P_i$, and for any pair $(\srange,p)\in \rangeS\times P$ with $\Phi(\srange,p)=1$,
there is a unique $i\le u$ such that $(\srange,p) \in \rangeS_i\times P_i$. The \emph{size} of 
$\biclique_\Phi$, denoted by $|\biclique_\Phi|$, is defined to be $\sum_{i=1}^u \left(|\rangeS_i|+|P_i|\right)$. 
Given $\biclique_\Phi$, both off-line range-searching and point-enclosure problems can be solved in 
$O(|\biclique_\Phi|)$ time.
Moreover, $\biclique_\Phi$ can also be used to answer \emph{on-line} range searching or point enclosure queries (for ranges $\sigma$ in the prescribed set $\Sigma$ or for points $p$ in the prescribed set $P$): 
For a range $\sigma$, say, access all the bicliques $(\Sigma_i,P_i)$ such that $\sigma\in\Sigma_i$, and return $\sum_i w(P_i)$ as the answer.
A symmetric approach handles point enclosure queries.
We thus focus on computing $\biclique_\Phi$, which is useful for other problems as well---see below.
By what has just been said, the real off-line component of our algorithm is the construction of the biclique partition.

\paragraph{Related work.}

We refer the reader to the survey papers~\cite{Ag,AE99,Mat94} for a review of range-searching. The 
best-known data structures for semi-algebraic 
range searching can answer a query, on an input set of $m$ points in $\reals^d$, in $O^*(m^{1-1/d})$ time using $O(m)$ space, 
or in $O(\log m)$ time using $O^*(m^s)$ space, where $s$ is the parametric dimension of 
the query ranges~\cite{AMS,MP,AAEZ}. By combining these data structures, in a so-called 
space/query-time tradeoff, we obtain, for any choice of $\store\in [m,m^s]$, 
a data structure that answers semi-algebraic range queries (for ranges of parametric dimension $s$)
in $O^*((m/\store^{1/s})^{\tfrac{1-1/d}{1-1/s}})$ time per query, using $O^*(\store)$ space and preprocessing.
Hence, with a suitable choice of $\store$, the total time 
taken (including preprocessing cost) in answering $n$ semi-algebraic queries is 
$O^*(m^{\tfrac{1-1/d}{1-1/ds}}n^{\tfrac{1-1/s}{1-1/ds}}+m+n)$~\cite{AAEKS22}. 
Afshani and Chang~\cite{AC21,AC22} showed that any data structure of size $\store$ 
needs $\Omega^* ((n^s/\store)^{1/\rho})$ time in the worst case, where $\rho=(s^2+1)(s-1)$,
to answer a two-dimensional semi-algebraic range-reporting query (for ranges of parametric dimension $s$) in the pointer machine model.
They also showed that if $P$ is a set of $n$ random points in $\reals^d$, a query can be answered in 
$O^*((n^s/\store)^{\tfrac{1}{3s-4}})$ time.

The problem of representing a graph compactly using cliques or bicliques has been studied for at least four
decades~\cite{CES83,Tu84}. For an arbitrary graph with $n$ vertices (including certain geometric graphs),
the worst-case bound on the size of the smallest biclique partition (again, the size of the partition is the sum of the sizes of the vertex sets of its bicliques) is $\Theta(n^2/\log n)$~\cite{Tu84}.
However, significantly better bounds are known for many geometric graphs, where 
the vertices are geometric objects (such as points, disks, segments, etc.) and two 
vertices are connected by an edge if the corresponding objects satisfy a simple geometric relation 
(such as two objects intersect, or be within distance $r$, for some parameter $r$). For example, interval graphs
on $n$ intervals on the real line admit a biclique partition of size $O(n\log n)$, point-orthogonal-box-incidence
graphs in $\reals^d$ admit such a representation of size $O((m+n)\log^{O(1)}n)$, unit-disk and segment-intersection 
graphs have a representation of size $O^*(n^{4/3})$ \cite{AV00,KS97}, and point-hyperplane incidence graphs 
admit an $O^*((mn)^{1-1/d}+m+n)$ representation size~\cite{ApS07}.
Recently, there has been some work on bounding the size of biclique partitions of 
general semi-algebraic geometric graphs (whose vertices are points in $\reals^d$ and whose
edges are defined by a semi-algebraic predicate of constant complexity)~\cite{AAEKS22,Do19}. 
We note though that, as already mentioned, not all geometric graphs, even in the plane, admit a small-size bipartite clique
partition~\cite{AAAS94}. Biclique partitions (as well as ``biclique covers'') have been effectively applied to study
extremal properties of geometric graphs, such as the regularity lemma, 
Zarankiewicz's problem, etc.~\cite{Do19,FPSS17,FPS16,FPS15}.
Most algorithms for computing these biclique partitions are based on off-line range-searching techniques; 
see, e.g., \cite{AV00,AS96,KS97}, affirming the close relationship between incidence and range-searching problems. 

In addition, faster algorithms for some basic graph problems have been proposed 
using biclique partitions (their running time being faster than what one could have obtained by running them 
on an explicit representation of the graph)~\cite{AKS24,AV00,CCCK24,FM95}. For example, BFS/DFS can 
be implemented in $O(N)$ time~\cite{AKS24,AV00} and a maximum bipartite matching in an 
intersection graph can be computed in $O^*(N)$ time~\cite{CCCK24}, assuming that a biclique partition of size $N$ is given.
The applicability of biclique partitions, however, goes far beyond basic graph algorithms. For example, the 
multipole algorithms for the so-called $n$-body problem, developed in the 1980's, can be regarded as 
an application of biclique partition of the complete graph of a set of points, where each 
biclique is \emph{well-separated}. Building on, and 
extending, this idea, Callahan and Kosaraju~\cite{CK93,CK95} introduced the notion of 
\emph{well-separated pair decomposition} (WSPD), showed the existence of small-size WSPD for point sets in $\reals^d$, 
and applied such decompositions to develop faster algorithms for many geometric proximity problems. 
Biclique partitions of geometric graphs have also been extensively used for a variety of geometric optimization problems~\cite{AAS97,AS96,AV00,KS97,Mat91}.

\paragraph{Our results.}
The main result of this paper is stated in the following theorem. 

\begin{theorem} 
	\label{thm:semioff}
Let $P$ be a set of $m$ points in $\reals^2$, and let $\rangeS$ be a set of $n$ semi-algebraic
	regions in $\reals^2$ with parametric dimension $s$, for some constant $s>0$.
	Let $\Phi: \rangeS \times P \rightarrow \{0,1\}$ be the Boolean semi-algebraic predicate (of constant complexity)
        such that $\Phi(\srange,p)=1$ if and only if $p\in\srange$.
	A biclique partition of $\rangeS \mathop{\Phi} P$ of size
\[
O^*\left( m^{\tfrac{2s}{5s-4}}n^{\tfrac{5s-6}{5s-4}} + m^{2/3}n^{2/3} + m + n \right) 
\]
	can be computed within the same randomized expected time (up to a subpolynomial factor).
\end{theorem}

This immediately implies the following corollary:
\begin{corollary}
	\label{cor:semioff} 
	Let $P$ be a set of $m$ points in $\reals^2$, let $\rangeS$ be a set of $n$ semi-algebraic
	regions in $\reals^2$  with parametric dimension $s$, for some constant $s>0$,
	let $(S,+)$ be a semigroup, and let $w:P\rightarrow S$ be a weight function. 
	The weights $w(\srange\cap P)$, for every $\srange\in \rangeS$, can be computed in 
	$O^*\left( m^{\tfrac{2s}{5s-4}}n^{\tfrac{5s-6}{5s-4}} + m^{2/3}n^{2/3} + m + n \right)$
	randomized expected time. Conversely, given a weight function $\eta:\rangeS \rightarrow S$, the weights
	$\sum_{\srange\in\rangeS:\srange\ni p} \eta(\srange)$, for every $p \in P$, can be computed within the same asymptotic time bound.
\end{corollary}

Our main observation is that the boundary arcs of the regions in $\rangeS$ can be processed to yield  
a family $\trapS$ of $O^*(n^{3/2})$ pseudo-trapezoids (or trapezoids for short), each bounded by (up to) two vertical lines and two subarcs of 
boundaries of regions in $\rangeS$, such that the edges of the trapezoids in $\trapS$ are pseudo-segments, i.e.,
any pair of edges of these trapezoids intersect in at most one point. Using the duality transform for pseudo-lines, 
proposed by Agarwal and Sharir~\cite{AS05}, we first present (in Section~\ref{sec:dual}) an algorithm for computing a biclique partition of $\trapS\Phi P$ , i.e., the set $\{ (\tau,p) \mid \tau\in \trapS, \; p\in P, \; p\in\tau\}$,
of size $O^*(m\sqrt{|\trapS|} + |\trapS|)$.
Using a standard hierarchical-cutting based 
method~\cite{AS05}, we improve (in Section~\ref{sec:primal}) the size of the biclique partition to $O^*(m^{2/3}n^{2/3}+n^{3/2})$, or even further
to $O^*(m^{2/3}\chi^{1/3}+n^{3/2})$, where $\chi$ is the number of intersections between the boundary curves. 
Finally, by working in the $s$-dimensional parametric space of $\rangeS$, we further improve the bound on the size 
of the biclique partition to
$O^*( m^{\frac{2s}{5s-4}}n^{\frac{5s-6}{5s-4}} + m^{2/3}n^{2/3} + m + n)$ (Section~\ref{sec:qspace}).

We conclude the discussion on our contributions by mentioning a few further applications of our results.
The off-line semi-algebraic range-searching problem arises in many different settings, as already reviewed earlier. Here we give one such 
example: Given a set $P$ of $m$ points in $\reals^2$ and a set $\rangeS$ of $n$ semi-algebraic regions (of constant complexity),
compute the smallest subset of $P$ that intersects all the regions in $\rangeS$ (the smallest \emph{hitting set}), or 
compute the smallest subset $\C$ of $\rangeS$ such that $P \subset \bigcup\C$ (the smallest \emph{set cover}).
Using the Br\"{o}nniman-Goodrich algorithm~\cite{BG95} for either of these problems, we can obtain an 
$O(\log \textsc{opt})$-approximate solution, where \textsc{opt} is the size of an optimal solution.
Each step of the algorithm in \cite{BG95} performs the following test:
given a set $\XX\subseteq P$ of points and a set $\C\subseteq\rangeS$ of geometric regions, 
determine whether $\srange\cap \XX\ne\emptyset$ for every region $\srange\in\C$, or 
test whether $p\in\bigcup\C$ for every $p\in \XX$. Our range-searching algorithm can be used to obtain 
a faster implementation of their algorithm.

As another application, our biclique-partition algorithm leads to faster implementation of basic graph algorithms
for geometric proximity graphs:
Let $P$ be a set $m$ of points in $\reals^2$, and let $\delta: \reals^2\times\reals^2 \rightarrow \reals_{\ge 0}$
be a \emph{semi-algebraic metric}, i.e., the unit disk $D_\delta = \{\xx \mid \delta(\xx,0) \le 1\}$ 
under $\delta(\cdot,\cdot)$ is a semi-algebraic set of 
constant complexity; $\delta$ is a metric when $D_\delta$ is a centrally symmetric convex set,
a convex distance function when $D_\delta$ is only convex, and just a distance function in general.
For a parameter $r\ge 0$, we can define a proximity graph $G_r(P) =(P,E)$, where 
$E=\{(p,q) \mid \delta(p,q) \le r\}$. A biclique partition of $G_r(P)$ can be computed using our algorithm, 
and its size depends on the parametric dimension of $\delta$.
As mentioned above, basic graph algorithms such as BFS and DFS on $G_r(S)$ can be implemented in time 
linear in the biclique partition size, so our result immediately yields a faster BFS/DFS 
algorithm for $G_r(S)$ (faster than what earlier methods yield).
Cabello~\etal~\cite{CCCK24} described an algorithm for computing the
maximum-size matching in a bipartite geometric-intersection graph, using a biclique partition. Combining their 
algorithm with
ours, one can obtain a faster algorithm for computing the minimum bottleneck matching between 
two point sets in $\reals^2$ under any semi-algebraic metric or distance function. 

\section{Bicliques Using Pseudo-Line Duality: The First Step}
\label{sec:dual}

Let $\trapS$ be a set of $n$ pseudo-trapezoids in $\reals^2$, each bounded from above and below by 
$x$-monotone semi-algebraic arcs with parametric dimension $s>0$, for some constant $s>0$, 
and from left and right by two vertical edges (some of these boundary arcs and edges may be absent). 
Furthermore, we assume that each pair of these arcs intersect in at most one point, i.e., the upper and
lower edges of the pseudo-trapezoids in $\trapS$ form a collection of \emph{pseudo-segments}. Let $P$ be a set of $m$ points in $\reals^2$.
Let $\trapS\mathop{\Phi} P \subseteq \trapS \times P$ be the set of pairs $(\trap, p)$ such that $p\in\trap$. The 
main result of this section is a randomized algorithm, with $O^*(m\sqrt{n}+n)$ expected running time, that 
constructs a biclique partition $\B := \biclique_\Phi(\trapS,P)$ of 
$\trapS\Phi P$ of size $O((m\sqrt{n}+n)\log^3 n)$.
We first give an overview of the algorithm, then describe 
its main steps in detail, and finally analyze its performance. This algorithm serves as the innermost routine
in our overall algorithm.

\subsection{Overview of the algorithm}
\label{subsec:overview}

We begin by defining two Boolean predicates 
$\Phi^\uparrow, \Phi^\downarrow: \trapS\times P \rightarrow \{0,1\}$ such
that $\Phi^\uparrow(\trap, p)=1$ (resp., $\Phi^\downarrow(\trap,p)=1$) if $p$ lies vertically above (resp., below)
the bottom (resp., top) arc of $\trap$. Note that $\Phi(\trap,p) = \Phi^\uparrow (\trap,p) \wedge \Phi^\downarrow (\trap,p)$.

The algorithm consists of the following high-level steps:
\medskip
\begin{description}
	\item[(i)] We construct a segment tree $T$ on the $x$-projections of the pseudo-trapezoids in $\trapS$. 
		Each node $v$ of $T$ is associated with an $x$-interval $I_v$ and the corresponding
		vertical slab $\slab_v = I_v \times \reals$. 
		A pseudo-trapezoid $\psi\in\trapS$ is stored at $v$ if the $x$-projection of $\psi$ 
		contains $I_v$ but does not contain $I_{\parent(v)}$, where $\parent(v)$ is the parent of $v$.
		Let $\trapS_v\subseteq \trapS$ be the set of pseudo-trapezoids stored at $v$, 
		clipped to within $\slab_v$, and let $P_v = P\cap\slab_v$. Set $n_v := |\trapS_v|$ 
		and $m_v := |P_v|$. By standard properties of segment trees, $\sum_v n_v = O(n\log n)$ and $\sum_v m_v = O(m\log n)$.

	\item[(ii)] 
		For each node $v$ of $T$, we compute a biclique partition 
		$\biclique_v := \biclique_\Phi(\trapS_v, P_v)$ of $\trapS_v \Phi P_v$, as follows.
		We partition $P_v$ into $r_v=\ceil{m_v/\sqrt{n_v}}$ subsets 
		$P_v^{(1)}, \ldots,  P_v^{(r_v)}$ of size at most $\sqrt{n_v}$ each. Set 
		$m_{v,i} := |P_v^{(i)}| \le \sqrt{n_v}$.
		We compute a biclique partition $\biclique_{v,i} := \biclique (\trapS_v, P_v^i)$ 
		for every $i \le r_v$, in (the following) two stages. 
		\begin{description}
			\item[(ii.a)] For every node $v\in T$ and for every $i \le r_v$, we compute a 
				biclique partition 
				$\biclique_{v,i}^\uparrow := \biclique_{\Phi^\uparrow} (\trapS_v, P_v^i)$.
			\item[(ii.b)] Next, for each biclique $(\trapS_j, P_j) \in \biclique_{v,i}^\uparrow$, 
				we compute a biclique partition 
				$\biclique_{v,i,j} := \biclique_{\Phi^\downarrow}(\trapS_j, P_j)$ of 
				$\trapS_j\Phi^\downarrow P_j$. We set 
				$\biclique_{v,i} = \bigcup_{(\trapS_j,P_j)\in \biclique_{v,i}^\uparrow} \biclique_{v,i,j}$.
		\end{description}
	\item[(iii)] We set $\biclique_v = \bigcup_{i=1}^{r_v} \biclique_{v,i}$  and return
		$\biclique = \bigcup_{v\in T} \biclique_v$ as the desired
		biclique partition $\biclique_\Phi(\trapS, P)$ (in which each clipped pseudo-trapezoid is replaced by
  the original pseudo-trapezoid containing it).
\end{description}

Steps~(ii.a) and (ii.b) are the  only nontrivial steps in the above algorithm. We describe the algorithm for 
Step~(ii.a). A symmetric procedure can be used for Step~(ii.b).

\subsection{Biclique partition for $\Phi^\uparrow$}
\label{subsec:arcs-clique}

Let $\slab$ be a vertical slab. Let $\Gamma$ be a set of $n$ $x$-monotone semi-algebraic arcs of constant complexity
whose endpoints lie on the boundary lines of $\slab$, so that any pair of arcs in $\Gamma$ intersect at most once, 
i.e., $\Gamma$ is a set of pseudo-segments. Let $P \subset W$ be a set of $m$ points. 
Slightly abusing the preceding notation, let 
$\Phi^\uparrow: \Gamma \times P \rightarrow \{0,1\}$ be a Boolean predicate such that 
$\Phi^\uparrow (\gamma, p) = 1$ if $p$ lies above $\gamma$ and $0$ otherwise. We describe a randomized
algorithm, with expected running time $O^*(m^2+n)$,
for computing a biclique partition $\biclique$ of $\Gamma\Phi^\uparrow P$ of size $O(m^2+n\log n)$.  
By choosing $P$ to be $P_v^i$ and $\Gamma$ to be the set of bottom arcs of the trapezoids in $\trapS_v$, we 
compute $\biclique_{\Phi^\uparrow} (\trapS_v, P_v^i)$, as required in Step~(ii.a).

Our algorithm consists of two stages. First, we rely on the pseudo-line duality transform 
described by Agarwal and Sharir~\cite{AS05}, as a major tool for the construction of the desired
biclique partition (see also~\cite{Go80}). The duality transform maps the arcs in $\Gamma$ to a 
set $\Gamma^*$ of dual points lying on the $x$-axis, and the points in $P$ to a set $P^*$ of dual 
$x$-monotone curves, such that $p$ lies above (resp., on, below) $\gamma$ 
if and only if the dual curve $p^*$ passes above (resp., through, below) the dual point $\gamma^*$. Furthermore, $P^*$ is a set of 
pseudo-lines, i.e., each pair of them intersect at most once. Agarwal and Sharir describe an $O^*(m^2+n)$-time 
sweep-line algorithm to construct $P^*$ and to compute a DCEL representation \cite{dBCKO} of the arrangement 
$\A(P^*)$, as well as the subset 
$\Gamma_f^* \subset\Gamma^*$ of dual points lying in each face $f$ (that meets the $x$-axis) of $\A(P^*)$.
Let $\gamma_1, \ldots, \gamma_n$ be the ordering of the arcs in $\Gamma$ in increasing order of the $y$-coordinates of their left endpoints. Then
the $x$-coordinate of the dual point $\gamma_i^*$ is $i$, for each $i$. Conversely, the dual curves are ordered 
in the $(+y)$-direction at $x=-\infty$, in the decreasing order of the $x$-coordinates of the primal points; 
see~\cite{AS05}. We note that the 
curves in $P^*$ do not have  constant combinatorial (or geometric) complexity, as each of them may contain
many breakpoints and turns, through which it weaves its way above and below the dual points of $\Gamma^*$ on the $x$-axis.
Nevertheless, we never need an explicit representation of a dual curve. The representation computed by the
algorithm in \cite{AS05} enables us to compute (i) the vertical ordering of a pair of curves at any given 
$x$-coordinate, and (ii) the (unique) intersection point between any pair of curves, in $O(1)$ time.

Second, we use geometric cuttings on $P^*$, the set of dual curves, to compute the desired bicliques.
More generally, let $\XX$ be a set of $m$ $x$-monotone arcs  in $\reals^2$ that are pseudo-segments,
let $\Delta$ be a pseudo-trapezoid such that it is either unbounded from its top/bottom or its 
top/bottom  edge is a portion of an arc of $\XX$,
and let $\chi$ be the number of vertices of $\A(\XX)$ inside $\Delta$. 
For a parameter $r>1$, a partition of $\Delta$ into 
a family $\Xi$ of pseudo-trapezoids, referred to as \emph{cells}, to distinguish them from the
input pseudo-trapezoids, is called a \emph{$(1/r)$-cutting} of $\XX$ within (or with respect to) 
$\Delta$ if every cell of $\Xi$ is crossed by at most $m/r$ arcs of $\XX$. 
(For $r>m$, cells of $\Xi$ are not crossed by an arc of $\XX$, i.e., $\Xi$ is a refinement of $\A(\XX)$.)
The \emph{conflict list} of a 
cell $\cell \in \Xi$, denoted by $\XX_\cell$, is the subset of arcs that cross $\cell$. 
We follow a hierarchical-cutting algorithm (as in~\cite{dBS,Ch93,Mat91}) to construct a $(1/r)$-cutting $\Xi$ 
of $\XX$ within $\Delta$. That is, we choose a sufficiently large constant $r_0$ and set $\nu = \ceil{\log_{r_0} r}$. 
We construct a sequence of cuttings $\cuthierarchy = \langle \Xi_0=\Delta, \Xi_1, \ldots, \Xi_\nu \rangle$ where $\Xi_i$ is a 
$(1/r_0^i)$-cutting of $\XX$ within $\Delta$, so the final cutting $\Xi_\nu$ is a $(1/r)$-cutting. 
$\Xi_i$ is obtained from $\Xi_{i-1}$ by computing for each 
cell $\cell\in\Xi_{i-1}$ a $(1/r_0)$-cutting of $\A(\XX_\cell)$ within $\cell$. 
(The construction in~\cite{dBS} ensures that the top and bottom edges of a cell in $\Xi_\cell$ is 
either a portion of an edge of $\cell$ or an arc of $\XX_\cell$.) 
Following the argument in~\cite{dBS}, 
it can be shown that the size of the $(1/r_0)$-cutting of $\XX_\cell$ within $\cell$ 
is at most $c_1(r_0+{\chi}_\cell r_0^2/m_\cell^2)$, where $m_\cell=|\XX_\cell|$, $\chi_\cell$ is the number
of vertices of $\A(\XX_\cell)$ within $\tau$, and $c_1$ is a constant independent of $r_0$.
Summing the bound over all cells $\cell$ of $\Xi_{i-1}$, using the fact that $\sum_{\cell\in\Xi_{i-1}} \chi_\cell = \chi$, and using an inductive argument (see, e.g.,~\cite{Ch93}), the size of $\Xi_i$ can be shown to be bounded by $c_1((c_2r_0)^i + r_0^{2i}\chi/m^2)$, where $c_2$ is some suitable constant independent of $r_0$ and $r$.
Therefore
$|\Xi_\nu| = O(r^{1+\eps}+\chi r^2/m^2)$, for any $\eps>0$, or $O^*(r) + O(\chi r^2/m^2)$ in our notation, provided $r_0$ is chosen sufficiently large.
In fact, the stronger bound $\sum_i |\Xi_i| = O^*(r) + O(\chi r^2/m^2)$ also holds.
Assuming that various primitive operations on the arcs of $\XX$ can be computed in $O(1)$ time,
the expected run time of this construction is $O(m^{1+\eps}+\chi r/m)$, for any $\eps>0$, which again we write as $O^*(m) + O(\chi r/m)$~\cite{Ch93}; see also~\cite{AS05}. 

In our context, after having computed $\A(P^*)$ as described above, a $(1/r)$-cutting of $P^*$ can be 
computed in $O^*(m+\chi r/m)$ time. We actually construct this cutting for $r=m+1$ (in this section only), so we get
a hierarchical $(1/(m+1))$-cutting $\cuthierarchy =\langle \Xi_0=\reals^2, \Xi_1, \ldots,  \Xi_\nu \rangle$ of $P^*$ in 
the dual plane, where $\nu=\ceil{\log_{r_0} (m+1)}$. Since $\chi=O(m^2)$, the size of the cutting is $O(m^2)$ and 
it can be computed in expected time $O^*(m^2)$.
Nevertheless, the more general setup considered above will be useful in another construction of a cutting, in the primal plane, which will be used in Section~\ref{sec:primal}.

In fact, for each $i$, the size of $\Xi_i$ is $O(r_0^{2i})$.
Since $r>m$, each cell of the final cutting $\Xi_\nu$ 
is not crossed by any arc of $P^*$. For every $i < \nu$ and for every cell 
$\cell\in \Xi_i$, let $P_\cell^* \subset P^*$ be the conflict list of $\cell$, and let $P_\cell=\{p \mid p^* \in P_\cell^*\}$. 
Let $\cell'\in \Xi_{i-1}$ be the parent cell that contains $\cell$. 
We associate a \emph{canonical subset} $P_\cell^\circ \subseteq P_{\cell'}$ with $\cell$, which is 
the set of points whose dual curves appear in the conflict list of its parent cell $\cell'$
and lie \emph{above} $\cell$ (without intersecting it), i.e.,
\[ 
P_\cell^\circ = \{ p_i \mid p_i^* \in P_{\cell'}^* \mbox{ and $p_i^*$ lies above $\cell$} \} .
\]
For $\cell\in\Xi_i$, $|P_\cell^\circ| \le m/r_0^{i-1}$.
Using the information computed by the Agarwal-Sharir algorithm~\cite{AS05}, we can check in $O(1)$ time, for each curve $p_i^*\in P_{\cell'}^*$, 
whether $p_i^*$ lies above $\cell$, and thereby obtain $P_\cell^\circ$.

Next, for a cell $\cell\in\Xi_i$, we set $\Gamma_\cell = \{ \gamma \in \Gamma \mid \gamma^* \in \cell\}$
(only cells that cross the $x$-axis are relevant). 
For every $i \le \nu$, $\sum_{\cell \in \Xi_i} |\Gamma_\cell| = n$.
We compute $\Gamma_\cell$ in a top-down manner. Suppose we have computed $\Gamma_\cell$ for a cell 
$\cell \in \Xi_i$.
For every (dual) point $\gamma^* \in \Gamma_\cell^*$, we compute which of the $O(r_0^2)=O(1)$ children cells of $\cell$
(in $\Xi_{i+1}$) contains $\gamma^*$. This step requires testing whether $\gamma^*$ 
lies inside a child cell $\hat\cell$ of $\cell$, which we can do in $O(1)$ time, as follows.
We can easily determine in $O(1)$ time whether $\gamma^*$ lies to the left (resp., to the right) of the right (resp., left) vertical edge of 
$\hat\cell$, but the top/bottom edge of 
$\hat\cell$ may have large complexity (due to the ``erratic'' way in which the dual arrangement is constructed in \cite{AS05}).
However, the top (or bottom) arc is a portion of a dual curve $p_i^*$, and the duality transform ensures that
$\gamma^*$ lies below/above $p_i^*$ if and  only if $\gamma$ lies below/above $p_i$. 
Since $\gamma$ is a semi-algebraic arc of constant complexity, we can test the above/below relationship between 
$p_i$ and $\gamma$ in $O(1)$ time.  Hence, we can distribute $\Gamma_\cell$ among its children cells in $O(|\Gamma_\cell|)$ time.
Summing over all levels of the hierarchy, the overall time spent in 
distributing the points of $\Gamma^*$ to the cells of $\cuthierarchy$ is $O(n \log m)$.

Finally,
we return 
\[
	\biclique_{\Phi^\uparrow} := \{(\Gamma_\cell, P_\cell^\circ) \mid \cell\in\Xi_i, 1 \le i \le \nu\}
\]
as the desired biclique partition of $\Gamma\Phi^\uparrow P$. 
\begin{lemma}
\label{lem:correctness}
	$\biclique_{\Phi^\uparrow}$ is indeed a biclique partition of $\Gamma\Phi^\uparrow P$.
\end{lemma}
\begin{proof} 
	By construction and the property of the dual transform, it is clear that all points of 
	$P_\cell^\circ$ 
	lie above all the arcs in $\Gamma_\cell$, i.e., $\Phi^\uparrow(\gamma, p)=1$ for every pair $(\gamma,p) \in \Gamma_\cell\times P_\cell^\circ$.  
	Conversely, let $(\gamma,p) \in \Gamma\times P$ be a pair such that $p$ lies above $\gamma$. 

	Let $\cell_0=\reals^2$ be the only cell of $\Xi_0$ and let $\cell_\nu$ be the cell of $\Xi_\nu$ 
	that contains $\gamma^*$.
        Clearly, $p\in P_{\cell_0}$ and $p \not\in P_{\cell_\nu}$ because the cells 
	of $\Xi_\nu$ are not crossed by any arc of $P^*$.
	Let $\hat\cell$ be the cell in $\cuthierarchy$
	for which $p\in P_{\hat\cell}$ and the index $i$ of
 its cutting $\Xi_i$ is the largest; $\hat\cell$ is a non-leaf node and $p\in P_\cell$ for all 
	ancestor cells $\cell$ of $\hat\cell$. Let 
	$\sigma$ be the child cell of $\hat\cell$ that contains $\gamma^*$.
	Since $p\not\in P_\sigma$, $\gamma^*\in\sigma$, 
	and $p^*$ lies above $\gamma^*$, we conclude that $p^*$ lies above $\sigma$ and thus $p\in P_\sigma^\circ$. 
	Hence, $(\gamma, p) \in \Gamma_\sigma \times P^\circ_\sigma$. Furthermore, $\sigma$ is the 
	only cell that contains $\gamma^*$ for which $p\in P_\sigma^\circ$. Therefore there is 
	a unique biclique in $\biclique_{\Phi^\uparrow}$ that contains the pair $(\gamma,p)$, 
	implying that  $\biclique_{\Phi^\uparrow}$ 
	is a biclique partition of $\Gamma\Phi^\uparrow P$, as claimed.
\end{proof}

We now bound the size of $\biclique_{\Phi^\uparrow}$ and the expected running time of the algorithm.
Recall that, for $1 \le i \le \nu$, we have $|\Xi_i|=O(r_0^{2i})$, $\sum_{\cell\in\Xi_i} |\Gamma_\cell| \le n$, and for any $\cell \in \Xi_i$, $|P^*_\cell| \le m/r_0^i$.
Therefore, the total size of $\biclique_{\Phi^\uparrow}$ is 
\begin{align*}
	\label{eq:above-size}
	|\biclique_{\Phi^\uparrow}| &= 
	\sum_{i=1}^\nu\sum_{\cell\in \Xi_i} O(|P_\cell| + |\Gamma_\cell|)
	= \sum_{i=1}^\nu O\biggl ( r_0^{2i} \cdot \frac{m}{r_0^{i}} + n\biggr )\\
	&= O\biggl (m\sum_{i=1}^\nu r_0^{i} + n \nu \biggr ) =  O(m^2+n\log m).
\end{align*}

Using similar considerations, the total expected time spent in computing $\biclique_{\Phi^\uparrow}$ is easily seen to be $O^*(m^2+n)$. 
Hence, we obtain the following result.

\begin{lemma} 
	\label{lem:above-below}
	Let $\Gamma$ be a set of $n$ $x$-monotone semi-algebraic arcs in $\reals^2$ of constant complexity, whose 
	endpoints lie on the boundary lines of a vertical slab $\slab$, and any pair of arcs in $\Gamma$ 
	intersect in at most one point, i.e., $\Gamma$ is a set of pseudo-segments. 
	Let $P \subset W$ be a set of $m$ points. Then a biclique partition of $\Gamma \mathop{\Phi^\uparrow} P$ 
	of size $O(m^2+n\log m)$ can be computed in expected time $O^*(m^2+n)$.
\end{lemma}

\subsection{Putting it all together}
\label{subsec:dual-plane}

Returning to the problem of computing a biclique partition of $\trapS \mathop{\Phi} P$, let $v$ be a node of the 
segment tree $T$, and let $\trapS_v$ and $P_v^{(1)}, \ldots, P_v^{(r_v)}$, $r_v = \ceil{m_v/\sqrt{n_v}}$
be the sets of pseudo-trapezoids and points, as defined above.  Set $n_v := |\trapS_v|$, $m_{v,i} := |P_{v,i}|$, and $m_v := |P_v|$.
For a pseudo-trapezoid $\trap_a \in \trapS_v$, let $\gamma_a^-, \gamma_a^+$ be its respective bottom and top boundary arcs. 
By construction, the endpoints of $\gamma_a^-, \gamma_a^+$ lie on the boundary lines of the vertical slab $\slab_v$,
so $\trap_a$ straddles $\slab_v$. Let $\Gamma_v^- = \{ \gamma_a^- \mid \trap_a \in \trapS_v\}$ be the set of 
bottom arcs of the pseudo-trapezoids in $\trapS_v$.
Fix a value $1\le i \le r_v$. We first compute a biclique partition $\biclique_{v,i}^\uparrow$ 
of $\Gamma_v^- \mathop{\Phi^\uparrow} P_v^{(i)}$ using the above algorithm. Let $\cuthierarchy_{v,i} = \{ \Xi_0, \ldots, \Xi_\nu\}$, $\nu=\ceil{\log_{r_0} m_{v,i}}$,
be the hierarchical cutting constructed by the algorithm, for task (ii.a) for the dual set of $P_v^{(i)}$.
Let $(\Gamma_\cell^-, P_\cell)$ be a biclique in this 
partition for some cell $\cell$ of a cutting in some $\cuthierarchy_{v,i}$, and let $\Gamma_\cell^+$ be the set of top arcs of the pseudo-trapezoids whose bottom arcs are 
in $\Gamma_\cell^-$, i.e., $\Gamma_\cell^+ = \{ \gamma_a^+ \mid \gamma_a^- \in \Gamma_\cell^-\}$. Following the same 
algorithm (but reversing the direction of the $y$-axis, so that it now solves an instance of type (ii.b)), we compute a biclique partition 
$\biclique_{v,i,\cell}$ of $\Gamma_\cell^+\mathop{\Phi^\downarrow} P_\cell$.  
For each resulting biclique $(\Gamma_{\cell,t}^+, P_{\cell,t}) \in \biclique_{v,i,\cell}$, 
we replace $\Gamma_{\cell,t}^+$ with $\trapS_{\cell,t} \subseteq \trapS$, which is 
the set of (the original input) trapezoids whose top arcs are in $\Gamma_\cell^+$. 
Abusing the notation a little, let $\biclique_{v,i,\cell}$ denote the resulting biclique partition.
We repeat this step for all bicliques $(\Gamma_\cell^-,P_\cell)$ 
in $\biclique(\Gamma_v^-,P_v^{(i)})$, set
$\biclique_{v,i} = \bigcup_{(\Gamma_\cell^-,P_\cell)\in \biclique_{v,i}^\uparrow} \biclique_{v,i,\cell}$, and return
$\biclique_{v,i}$ as a biclique partition of $\trapS_v \mathop{\Phi} P_v^{(i)}$. 
By repeating this step for all $i\le r_v$ and for all nodes $v\in T$, we obtain the desired biclique 
partition $\biclique := \biclique_\Phi(\trapS,P)=\bigcup_{v\in T}\bigcup_{i=1}^{r_v} \biclique_{v,i}$. 
It is easy to check that, by construction and the properties of segment trees, the resulting collection of 
bicliques is edge disjoint, and its union gives all pairs $(p,\srange)$ with $p\in\srange$, so it is indeed a 
biclique partition of the desired form. It remains to bound the size of $\biclique$.

Consider a cutting $\Xi_j$ in $\cuthierarchy_{v,i}$, as constructed above, for some parameters $v$, $i$, and let $\cell$ be a cell in $\Xi_j$. Let $(\Gamma_\cell^-, P_\cell)$ be the biclique in $\Gamma_v^-\Phi^\uparrow P_v^{(i)}$ 
corresponding to $\cell$. 
By Lemma~\ref{lem:above-below}, 
\[ 
|\biclique_{v,i,\cell}| = O(|P_\cell|^2 + |\Gamma_\cell^-| \log |P_\cell|).
\]
Furthermore,  
$|\Xi_j| = O(r_0^{2j})$, 
$|P_\cell| = O(m_{v,i}/r_0^{j})$, and
$\sum_{\cell \in \Xi_j} |\Gamma_\cell^-|= n_v$.
Hence, summing over all cells of $\Xi_j$ and over all cuttings $\Xi_j$ in $\cuthierarchy_{v,i}$, the size of 
$\biclique_{v,i}$ is 
\begin{align*}
	|\biclique_{v,i}| &= \sum_{j=1}^\nu \sum_{\cell\in \Xi_j} 
				O(|P_\cell|^2 + |\Gamma_\cell^-| \log |P_\cell|) \\
			&= \sum_{j=1}^\nu O\left ( r_0^{2j}\cdot \frac{m_{v,i}^2}{r_0^{2j}} + 
					n_v\log m_{v,i} \right )
			 = \sum_{j=1}^\nu O(m_{v,i}^2 + n_v\log m_{v,i})\\
			&= O(m_{v,i}^2 \log m_{v,i} + n_v\log^2 m_{v,i})
			= O(n_v\log^2 n_v) ,
\end{align*}
because $m_{v,i} \le \sqrt{n_v}$. 
Summing over all $i \le r_v=\ceil{\tfrac{m_v}{\sqrt{n_v}}}$, the size of $\biclique_v$ is 
$O((m_v\sqrt{n_v}+n_v)\log^2 n_v)$. That is, we have shown: 
\begin{lemma} 
	\label{lem:primal1} 
	Let $\trapS$ be a set of $n$ pseudo-trapezoids in $\reals^2$, each bounded from above and below 
	by $x$-monotone semi-algebraic arcs of constant complexity, such that any pair of these arcs 
	intersect in at most one point, so that the vertical edges of the trapezoids lie on the boundary lines of some 
	vertical slab $W$.
	Let $P$ be a set of $m$ points lying in $W$. Then a biclique partition of 
	$\trapS \mathop{\Phi} P$ of size $O((m\sqrt{n}+n)\log^2 n)$ can be computed in 
	expected time $O^*(m\sqrt{n}+n)$.
\end{lemma}

Finally, summing the size of the biclique partitions over all nodes $v$ of the segment tree 
$T$ and plugging the values $\sum_{v\in T} m_v = O(m\log n)$, $\sum_{v\in T} n_v= O(n\log n)$,
we obtain the following summary result of this section:
\begin{corollary} 
	\label{cor:primal1} 
	Let $\trapS$ be a set of $n$ pseudo-trapezoids in $\reals^2$, each bounded from above and below 
	by $x$-monotone semi-algebraic arcs of constant complexity, such that any pair of these arcs 
	intersect in at most one point, and
	let $P$ be a set of $m$ points in $\reals^2$. Then a biclique partition of 
	$\trapS \mathop{\Phi} P$ of size $O((m\sqrt{n}+n)\log^3 n)$ can be computed in 
	expected time $O^*(m\sqrt{n}+n)$.
\end{corollary}

\section{Bicliques Using Cuttings in the Primal: The Second Step}
\label{sec:primal}

Let $P$ be a set of $m$ points and $\rangeS$ a set of $n$ semi-algebraic sets
of constant complexity in the plane, as defined in the introduction. 
Our goal is to compute a biclique partition $\biclique_\Phi (\rangeS, P)$ for the inclusion predicate $\Phi$, i.e., 
$\Phi(\srange, p)=1$ iff $p\in\srange$.  Let $\Gamma$ denote the set of boundary edges of the regions in $\rangeS$, each of which is a semi-algebraic arc of constant complexity.
Without loss of generality, we assume that each of these arcs is $x$-monotone, because we can split every non-monotone arc into $O(1)$ $x$-monotone subarcs.
See below for further elaboration of this issue.

Following the technique in \cite{SZ} (see also~\cite{ArS}), we cut the arcs in $\Gamma$
into $O^*(n^{3/2})$ subarcs
that constitute a family of pseudo-segments, i.e., each pair of subarcs intersect at most
once. 
Agarwal~\etal~\cite{AAEZ} (see also \cite{AEZ}) present an efficient algorithm 
for constructing these cuts, which runs in $O^*(n^{3/2})$ 
randomized expected 
time.
This step partitions the edges of each region
$\srange \in \rangeS$ into subarcs, which we view as new edges of $\srange$. We compute the vertical decomposition 
of $\srange$, or rather of the collection of subarcs constituting its boundary. This divides $\srange$ into a set of pseudo-trapezoids and in general further partitions its edges into smaller
pieces. Each resulting pseudo-trapezoid is bounded by at most two vertical edges and two  
(top and bottom) semi-algebraic arcs that are portions of the split subarcs of the edges of $\srange$.
Let $\trapS$ denote the resulting set of pseudo-trapezoids, and let $\Gamma$ denote the set of 
	their top and bottom edges. 
Set $N := |\trapS|$, so $|\Gamma|\le 2N$; by construction, $N= O^*(n^{3/2})$. Let $\chi=O(n^2)$ denote the number of intersection points between the 
arcs of $\Gamma$. 
It suffices to construct a biclique partition for $\trapS \mathop{\Phi} P$ (that is, for 
$\trapS$ instead of $\rangeS$) and then replace each trapezoid by its containing region, in each biclique. 
(In fact, the forthcoming algorithm will run on sets of smaller pseudo-trapezoids, each contained in some 
pseudo-trapezoid of $\Psi$, but the same replacement rule applies.)
The algorithm described in the previous section already computes such a biclique partition of size
$O((m\sqrt{N}+N)\log^3 N) = O^*(mn^{3/4}+n^{3/2})$, within the same expected time.  
In this section, we show how to improve the bound to 
$O^*(m^{2/3}\chi^{1/3}+n^{3/2}) = O^*(m^{2/3}n^{2/3}+n^{3/2})$, using hierarchical cuttings 
of $\Gamma$~\cite{AS05,Ch93}, as in the preceding section but in the primal plane.
This approach is analogous to the widely used approach for obtaining sharp bounds on 
various substructures of arrangements of curves in the plane or for the number of incidences between points
and curves in the plane (see, e.g.,~\cite{AS05,CEGSW90,SA95}).
Specifically, our analysis proceeds as follows.

We follow the same overall algorithm as described in Section~\ref{subsec:overview}, now in the primal plane, with a few suitable modifications.
(We borrow some notations from Section~\ref{sec:dual}, but remind the reader that we are now in the primal plane.)
Let $v$ be a node of the segment tree $T$, let $W_v$ be the vertical slab associated with $v$,
and let $\trapS_v, P_v$ be the subsets of pseudo-trapezoids and points stored at $v$, where the trapezoids of 
$\Psi_v$ are clipped to within $W_v$. Let $\Gamma_v$ be the set of top and bottom arcs in the pseudo-trapezoids of 
$\trapS_v$. Because of the clipping, the endpoints of the arcs of $\Gamma_v$, and thus the vertical edges of
the trapezoids of $\trapS_v$, lie on the boundary lines of $W_v$. Set $N_v = |\trapS_v|$, $m_v = |P_v|$, and set
$\chi_v$ to be the number of intersection points 
between the arcs of $\Gamma_v$. Here $\sum_v N_v = O(N \log n)$, $\sum_v m_v = O(m\log n)$, and $\sum_v \chi_v \le \chi = O(n^2)$.
We compute a biclique partition $\biclique_v$ of $\trapS_v\Phi P_v$, as follows.

Fix a parameter $r>1$, whose precise value will be set later. As described in 
Section~\ref{subsec:arcs-clique}, we choose $r_0$ to be a sufficiently large constant, set 
$\nu = \ceil{\log_{r_0} r}$, and 
construct a hierarchical $(1/r)$-cutting 
$\cuthierarchy = \langle \Xi_0= W_v, \Xi_1, \ldots, \Xi_\nu\rangle$ of $\Gamma_v$ (in the primal plane) of 
total size $O^*(r) + O(\chi_v r^2/N_v^2)$, in expected time $O^*(N_v +\chi_v r/N_v)$. More generally, 
for $1 \le i \le \nu$, $\Xi_i$ is a $(1/r_0^i)$-cutting of $\Gamma_v$ of size $O^*(r_0^i)+O(\chi_vr_0^{2i}/N_v^2)$.
Unlike the algorithm of the previous section, here we do not construct the cutting until the leaf subproblems are of constant size, but stop when we reach the target value $r$.
For every $i \le \nu$ and for every cell $\cell\in \Xi_i$, let $\trapS_\cell$ be the set of 
pseudo-trapezoids in $\Psi_v$ whose boundaries cross $\cell$. Let $\cell'\in \Xi_{i-1}$ be the parent cell that contains 
$\cell$. We set 
$\C_\cell = \{ \trap \in \trapS_{\cell'} \mid \cell\subseteq \trap\}$ to be the set of 
pseudo-trapezoids of $\trapS_{\cell'}$ that contain $\cell$. Set $P_\cell = P \cap \cell$,
$N_\cell=|\trapS_\cell|$, and $m_\cell=|P_\cell|$.
Finally, for each cell $\cell$ of the bottom cutting $\Xi_\nu$, we compute a biclique partition $\biclique_\cell$ of 
$\trapS_\cell \mathop{\Phi} P_\cell$ using the algorithm described in the previous section in the dual setting
(cf.\ Lemma~\ref{lem:primal1}). 
We set
\begin{equation}
	\label{eq:overall-clique}
	\biclique_v = \{ (\C_\cell,P_\cell) \mid \cell \in \Xi_i, 1 \le i \le \nu \} \cup
				\bigcup_{\cell \in \Xi_\nu} \biclique_\cell .
\end{equation}
We repeat this step for all nodes $v$ of the segment tree $T$ and return $\bigcup_{v\in T} \biclique_v$ as the desired 
biclique partition of $\trapS_v\Phi P_v$.
Following an argument similar to that in Lemma~\ref{lem:correctness}, we can argue that $\biclique_v$ 
is indeed a biclique partition of $\trapS_v\mathop{\Phi} P_v$ (i.e., its bicliques are edge disjoint and cover all edges of $\trapS_v\Phi P_v$).

We now analyze the size of $\biclique_v$ and the running time of the algorithm.
Since $r_0$ is a constant and we have already computed conflict lists for each cell $\cell$, we get that
$\trapS_\cell, \C_\cell, P_\cell$, for all cells $\cell$ over all cuttings, can be computed in 
$O^*(N_v) + O(m_v\log r + \chi_v r/N_v)$ expected time.
By Lemma~\ref{lem:primal1}, computing $\B_\cell$ takes $O^*(m_\cell N_\cell^{1/2} + N_\cell)$ expected 
time. Since $N_\cell \le  N_v/r$ and $\sum_\cell m_\cell = m_v$, the expected time spent in computing 
$\biclique_\cell$, over all cells $\cell$  of $\Xi_\nu$, is
\begin{align}
	\sum_{\cell\in\Xi_\nu} O^*(m_\cell N^{1/2}_\cell+N_\cell) &= 
	O^* \left ( \frac{N_v^{1/2}}{r^{1/2}}\sum_{\cell \in \Xi_\nu} m_\cell + \frac{N_v}{r} |\Xi_\nu| \right ) \nonumber\\
	&= O^* \left ( \frac{N_v^{1/2}}{r^{1/2}} m_v +  \chi_v \frac{r}{N_v} + N_v \right ) .
	\label{eq:primal-time}
\end{align}
We choose
$$
r = \min \left\{ N_v, \left\lceil{ N_v m_v^{2/3}/\chi_v^{2/3}}\right\rceil\right\}.
$$ 
Note that if $r=N_v$ then $\chi_v \le m_v$, so in this case the bound is $O^*(m_v + N_v)$.
Plugging this value of $r$ into (\ref{eq:primal-time}), the expected running time is
$O^*(m_v^{2/3}\chi_v^{1/3} + m_v + N_v)$. 
This also bounds, up to the $O^*$ notation, the size of $\bigcup_{\cell\in\Xi_\nu} |\biclique_\cell|$. 

To bound the size of the first term in 
(\ref{eq:overall-clique}), we observe that $\sum_\cell |P_\cell|$, where the sum is taken over all cells $\cell$ of all the cuttings in $\cuthierarchy$, is 
$O(m_v\log r) = O^*(m_v)$. Similarly, 
\[
\sum_{i=1}^\nu \sum_{\cell\in\Xi_i} |\C_\cell| = O^*(N_v) + O(\chi_v r/N_v) = O^*(m_v^{2/3}\chi_v^{1/3}+m_v+N_v).
\]
Hence, the total size of $\biclique_v$ is $O^*(m_v^{2/3}\chi_v^{1/3}+m_v+N_v)$.
The same bound, up to the $O^*$ notation, applies to the expected running time of the algorithm.

Summing the above bound over all nodes $v$ of $T$ and plugging the values $\sum_v m_v = O(m\log n)$,
$\sum_v N_v = O(N\log n)$, $\sum_v \chi_v \le \chi = O(n^2)$, and $N=O^*(n^{3/2})$, the expected running time, as well 
as the size of $\biclique$, are
$O^*(m^{2/3}n^{2/3}+m+n^{3/2})$.
Putting everything together, we obtain the following summary lemma of this section.

\begin{lemma} 
	\label{lem:primal} 
	Let $P$ be a set of $m$ points and $\rangeS$ a set of $n$ semi-algebraic sets 
	of constant complexity  in $\reals^2$.
	Then a biclique partition of $\rangeS \mathop{\Phi} P$ of size $O^*(m^{2/3}n^{2/3}+m+n^{3/2})$ can be 
	computed in expected time $O^*(m^{2/3}n^{2/3}+m+n^{3/2})$. If $\chi$ is the number of intersection 
	points between the edges of $\rangeS$, then the size and the expected running time reduce to 
	$O^*(m^{2/3}\chi^{1/3}+m+n^{3/2})$.

\end{lemma}

\section{Bicliques in Query Space: The Final Step}
\label{sec:qspace}

A weakness of the above algorithm is that the $n^{3/2}$ term in the bounds on the size and the running time
dominates for $m < n^{5/4}$. (A similar issue arises in earlier studies of combinatorial bounds; see, 
e.g.,~\cite{ANPPSS,SZ}.) To mitigate the effect of this term for such smaller values of $m$, we apply a 
divide-and-conquer technique in the $s$-dimensional parametric space of the query regions, 
which now become points, so that 
the number of query regions reduces more rapidly than the number of input points, which become surfaces,
in the recursive subproblems. 
When we reach subproblems for which $m\ge n^{5/4}$, we switch back to the two-dimensional plane and 
apply Lemma~\ref{lem:primal}. This process yields the improved bound promised in Theorem~\ref{thm:semioff}.  

For simplicity, we assume that the regions in $\rangeS$ are defined by a single polynomial inequality. 
Namely, there is an $(s+2)$-variate polynomial $g(\xx,\yy): \reals^2\times\reals^s \rightarrow \reals$ such 
that each $\srange_i\in\rangeS$ is of the form $g(\xx,\yy_i)\ge 0$ for some $\yy_i \in \reals^s$.
Extending this setup to the general case of semi-algebraic regions (with a more involved defining predicate) 
is not difficult, 
and will be discussed later.
We denote $\yy_i$ as $\tilde\srange_i$, which is a representation of $\srange_i$ as a point in $\reals^s$.
Set $\tilde\rangeS = \{ \tilde\srange_i \mid 1 \le i \le n\} \subset \reals^s$. For 
each $p_j \in P$, we define a semi-algebraic set 
$\tilde{p}_j = \{ \yy\in\reals^s  \mid g(p_j, \yy)\ge 0\}$, namely the set of points representing regions that contain $p_j$. 
Set $\tilde P = \{ \tilde p_j \mid 1 \le j \le m\}$. 
Clearly, $p_j \in \srange_i$ if and only if $\tilde\srange_i \in \tilde p_j$. 
Thus a biclique $(\tilde P_a, \tilde\rangeS_a)$ of $\tilde P \mathop{\Phi}\tilde\rangeS$ directly corresponds to a 
biclique $(\rangeS_a,P_a)$ of $\rangeS \mathop{\Phi} P$.

We use the polynomial-partitioning technique, initiated by Guth and Katz~\cite{GK15}, and made algorithmic later in
\cite{AAEZ,MP}, 
for computing bicliques of $\tilde P\mathop{\Phi}\tilde \rangeS$. In particular, we rely on the following 
result by Matou\v{s}ek and Pat\'akov\'a \cite{MP}, used for constructing a partition tree for on-line semi-algebraic range searching:

\begin{lemma}[Matou\v{s}ek and Pat\'akov\'a \cite{MP}]
  \label{lem:mp}
  Let $V$ be an algebraic variety of dimension $k\ge 1$ in $\reals^d$ such that all of its irreducible 
  components have dimension $k$ as well, and the degree of every polynomial defining $V$ is at most 
  some parameter $E$. 
  Let $S \subset V$ be a set of $n$ points, and let $D\gg E$ be a parameter. 
  There exists a polynomial $g\in\reals[x_1,\ldots,x_d]$ of degree at most 
  $E^{d^{O(1)}}D^{1/k}$ that does not vanish identically on 
  any of the irreducible components  of $V$ (i.e., $V\cap Z(g)$ has dimension at most $k-1$), 
  and each cell of $V\setminus Z(g)$ contains at most 	$n/D$ points of $S$.  Assuming $D, E, d$ are constants,
  the polynomial $g$, a semi-algebraic representation of each cell in  $V\setminus Z(g)$, and the points of 
  $S$ lying in each cell, can be computed in $O(n)$ time.
\end{lemma}

\subsection{Algorithm} \label{sec:salg}
\label{subsec:algo}

We now describe the algorithm for computing the biclique partition.
A seeming complication in using Lemma~\ref{lem:mp} is that it does not provide any guarantees on the 
partitioning of the points that lie on $Z(g)$.  As such, we have to handle $S\cap Z(g)$ separately. 
Nevertheless, the lemma does provide us with the means of doing this, as it is formulated in terms of point sets 
lying on a variety of any dimension. This leads to two different threads of recursion---one of them recurses on  
subproblems of smaller size, as in the earlier algorithms, and the other recurses on the dimension of the variety 
that contains the point set. We will view each recursive subproblem as associated with a node $v$ of the 
recursion tree, which will naturally be a multi-level structure (two main levels for now, but the number will grow
when we handle later more general ways of defining the regions in $\Sigma$). Each recursive subproblem, at some node $v$, consists of a 
triple $(\F_v, \rangeS_v, P_v)$, where $\F_v$ is a set of $O(1)$ $s$-variate polynomials of constant degree in
$\reals[\yy]$, and $\rangeS_v \subseteq \rangeS$ is a set of regions such that $\tilde\rangeS_v \subset Z(\F_v)$,
where $Z(\F_v) = \bigcap_{F\in\F_v} Z(F)$ is the common zero set of $\F_v$, and $P_v \subseteq P$. 
Initially, $\F_v=\emptyset$ and $Z(\F_v)=\reals^s$, $\rangeS_v=\rangeS$, and $P_v=P$. The goal is to 
compute a biclique partition $\biclique_v$ of $\rangeS_v\mathop{\Phi} P_v$, in a recursive manner.

Let $\sd_v$ denote the dimension of $Z(\F_v)$, and put $n_v = |\rangeS_v|$ and $m_v=|P_v|$. We stop the recursion
as soon as either $m_v > n_v^{5/4}$ or $n_v \le n_0$, for some constant parameter $n_0$ that we will set later.

We first consider the case $\sd_v=1$. For simplicity, assume that $Z(\F_v)$ is a connected curve (the general
case is handled by partitioning $Z(\F_v)$ into its connected components and handling each of them separately).
In this case, the points of $\tilde\rangeS_v$ lie on a one-dimensional connected curve. Furthermore, for any 
$p_i \in P_v$, $\tilde p_i \cap Z(\F_v)$ is a collection of $O(1)$ intervals. Therefore 
a biclique partition of $\rangeS_v\mathop{\Phi} P_v$ of size $O((m_v+n_v)\log n_v)$ can easily be computed 
using $1$-dimensional range trees~\cite{dBCKO}. 

Next, assume that $\sd_v > 1$. 
If $m_v > n_v^{5/4}$, we compute a biclique partition $\biclique_v$ of $\rangeS_v \mathop{\Phi} P_v$ using 
the algorithm described in Section~\ref{sec:primal} (cf.\ Lemma~\ref{lem:primal}), i.e., ignoring the dual
representation in $\reals^{\sd_v}$. The size of the partition is then $O^*(m_v^{2/3}n_v^{2/3} + m_v)$. If $m_v \le n_v^{5/4}$ and
$n_v \le n_0$, the problem is of constant size, and we can output any trivial biclique partition, say one consisting
of single-edge graphs. So assume that $m_v \le n_v^{5/4}$ and $n_v > n_0$.

We choose a sufficiently large constant $D :=  D(s_v)$ and apply Lemma~\ref{lem:mp}, which
yields a partitioning polynomial $F_v$ for the point set $\tilde\rangeS_v$, with respect to the variety $Z(\F_v)$,
that satisfies the properties of the lemma. The degree of $F_v$ is at most 
$E^{s^{O(1)}}D^{1/\sd_v}$,
where $E$ is the degree of $Z(\F_v)$. Fix $\eps > 0$ to be an arbitrarily small number. By choosing
$D = E^{2s^{O(1)}s_v/\eps}$ so that $E^{s^{O(1)}} = D^{\eps/2s_v}$, we make the degree of $F_v$ at 
most $D^{(1+\tfrac{\eps}{2})/s_v}$.
Moreover, $F_v$ does not vanish on any irreducible component of $Z(\F_v)$, and each cell 
(connected component) of $Z(\F_v)\setminus Z(F_v)$ contains at most $n_v/D$ points of $\tilde\rangeS_v$.
Let $\Xi$ be the set of cells of $Z(\F_v)\setminus Z(F_v)$. For every cell $\cell\in\Xi$, 
we define $\rangeS_\cell :=  \{ \srange\in\rangeS_v \mid \tilde\srange \in \cell\}$,
$P_\cell :=  \{ p\in P_v \mid \tilde p \cap \cell \ne \emptyset \;\wedge\; \cell \not\subset \tilde p \}$, and 
$P^\circ_\cell := \{ p\in P_v \mid \cell \subseteq \tilde p\}$. That is, $P_\cell$ (resp., $P^\circ_\cell$)
is the set of points whose dual regions cross $\cell$ (resp., fully contain $\cell$).
Since $\tilde\srange \in \cell \subset \tilde p$ for every pair $(\srange,p) \in \rangeS_\cell\times P_\cell^\circ$,
we add the pair $(\rangeS_\cell, P_\cell^\circ)$, as one of the bicliques, to $\biclique_v$. 
We recursively compute a biclique partition for the subproblem 
$(\F_v, \rangeS_\cell, P_\cell)$, and add all the resulting bicliques to $\biclique_v$. 

Finally, we need to cater to the remaining set
$\rangeS_{v,0} = \{ \srange \in \rangeS_v \mid \tilde\srange \in Z(F_v)\}$.  
Set $n_{v,0} = |\rangeS_{v,0}|$. We recursively compute (now recursing on the dimension of the containing variety) 
a biclique partition $\biclique_{v,F_v}$ for the subproblem $(\F_v\cup\{F_v\}, \rangeS_{v,0},P_v)$, and add its 
bicliques to $\biclique_v$. Note that the dimension of $Z(\F_v\cup\{F_v\}) = Z(\F_v)\cap Z(F_v)$ is at most $\sd_v-1$
(see Lemma~\ref{lem:mp}), so this indeed yields a recursion on the dimension.

We return the overall resulting collection $\biclique_v$ as the desired biclique partition of $\rangeS_v \mathop{\Phi} P_v$. The final output, at the root of the recursion, is the
desired biclique partition of $\rangeS \mathop{\Phi} P$.

\subsection{Analysis}

Using an inductive argument, it can be shown that the algorithm described above returns a biclique partition of $\rangeS_v \mathop{\Phi} P_v$, for each recursive node $v$, so, in particular, it yields a biclique partition of $\rangeS \mathop{\Phi} P$.
We now bound the size of the biclique partition computed by the algorithm. The same 
analysis will also bound the expected running time of the algorithm by the same bound (up to the $O^*(\cdot)$
notation). We need the following result from real algebraic geometry for our analysis:
\begin{lemma}[{Barone and Basu~\cite{BB12}}]
  \label{lem:BB}
  Let $V$ be a $k$-dimensional algebraic variety in $\reals^d$
  defined by a finite set $\G$ of $d$-variate polynomials, each of degree at most 
  $\Delta$, and
  let $\F$ be a set of $t$ polynomials of degree at most $\Delta'\ge\Delta$. Then the number of cells
  of $\A(\F\cup\G)$ (of all dimensions) that are contained in $V$ is bounded by $O(1)^d \Delta^{d-k}(t\Delta')^k$.
\end{lemma}

Let $\beta(n_v,m_v,s_v)$ denote the maximum size of the biclique partition returned by the above algorithm 
for a subproblem $(\F_v,\rangeS_v,P_v)$ where $Z(\F_v)$ has dimension $\sd_v$, $|\rangeS_v| \le n_v$, and 
$|P_v| \le m_v$. We derive a recurrence for $\beta(n_v,m_v,s_v)$. First, as mentioned above,
$\beta(n_v,m_v,1) = O((n_v+m_v)\log m_v)$. For $\sd_v>1$, 
if $n_v\le n_0$, we can output a trivial biclique partition, consisting of single-edge bicliques, of size $O(m_v)$, 
so we have $\beta(n_v,m_v,s_v) = O(m_v)$ in this case.
For $m_v \ge n_v^{5/4}$ we have
\[
\beta(n_v,m_v,s_v) = O^*(m_v^{2/3}n^{2/3}+m_v+n_v^{3/2}) =
O^*(m_v^{2/3}n^{2/3}+m_v)
\]
(cf.\ Lemma~\ref{lem:primal}).

It remains to consider the case $\sd_v>1$, $m_v < n_v^{5/4}$ and $n_v>n_0$. The size of $\biclique_v$ is the overall 
size of the biclique partitions returned by the recursive subproblems, plus the size of the nonrecursive part of the partition, where the latter size is bounded by
\[ 
\sum_{\cell\in\Xi} \left( |\rangeS_\cell| + |P_\cell^\circ| \right) = \sum_{\cell\in\Xi} (n_\cell + m_v) \le n_v + |\Xi| m_v.
\]
The size of the biclique partitions computed recursively on the cells of $\Xi$ is at most 
\[
\sum_{\cell\in\Xi} \beta(n_\cell,m_\cell,s_v),
\]
and the size of the partition computed for 
$\rangeS_{v,0}$ is at most $\beta(n_{v,0},m_v,s_v-1)$.

By Lemma~\ref{lem:BB}, with $\Delta = E$, $k = s_v$, $t =1$, and $\Delta' =D^{(1+\tfrac{\eps}{2})/s_v}$,
and by our choice of $D= E^{2s^{O(1)}s_v/\eps}$, we have
\[
|\Xi| = O\left( E^s (D^{(1+\tfrac{\eps}{2})/s_v})^{s_v} \right) \le c_4 D^{\eps/2s_v} \cdot D^{1+\eps/2} =
c_4 D^{1+\eps} ,
\]
for some constant $c_4>0$ that depends on $s$.
For a point $p\in P_v$, the number of cells of $\Xi$ crossed by $\tilde p$ is the number of cells in
$(Z(\tilde p)\cap Z(\F_v))\setminus Z(F_v)$, which, by Lemma~\ref{lem:BB}, is at most $c_4D^{(1+\eps)(s_v-1)/s_v}$ (note that in this case we use similar considerations as above with $k = s_v - 1$, and that the factor $\Delta^{d-k}$ in the lemma, which is now $E^{s - s_v + 1}$, becomes relatively negligible in terms of $D$).
Hence, $\sum_\cell m_\cell  \le c_4 m_v D^{(1+\eps)(1-1/s_v)}$.  By Lemma~\ref{lem:mp}, $n_\cell \le n_v/D$ for every 
$\cell\in\Xi$, and $\sum_{\cell \in \Xi} n_\cell + n_{v,0} \le n_v$. 

Putting everything together, we obtain the following recurrence for $\beta(n_v,m_v,s_v)$, where in the third case we have made the $O^*$ notation explicit, with the $\eps$ used above:
{\small
\begin{equation}
	\label{eq:recur}
	\!\! \beta(n_v,m_v,\sd_v)\!\! \le\!\! \left\{\!\! \begin{array}{ll}
		c_1 (m_v+n_v)\log m_v & \!\! s_v=1,\\[1mm]
        c_2 m_v & \!\!\! s_v>1, n_v \le n_0,\\[1mm]
		c_3(m_v^{2/3+\eps}n_v^{2/3}+m_v) & \!\! s_v>1, m_v \ge n_v^{5/4},\\[1mm]
		\displaystyle \sum_{\cell\in \Xi} \beta(n_\cell,m_\cell,s_v) + 
			\beta(n_{v,0},m_v,s_v-1)   + c_4(n_v + D^{1+\eps} m_v) &\!\!\!  s_v>1, m_v < n_v^{5/4}, n_v > n_0,
	\end{array}
	\right .
\end{equation}
}
where $c_1$ is some absolute constant, $c_2$ is a constant that depends on $n_0$, $\eps>0$ is an arbitrarily small
constant, $c_3$ is a constant that depends on $\eps$, and $c_4$ is as above.
Furthermore, $n_\cell \le n/D$ for all $\cell\in \Xi$, $\sum_{\cell\in\Xi} n_\cell + n_{v,0} = n_v$, and 
$\sum_{\cell\in\Xi} m_\cell \le c_4 m_v D^{(1+\eps)(1-1/s_v)}$. We claim that the solution to the recurrence (\ref{eq:recur}) is 
\begin{equation}
	\label{eq:rec-sol}
	\beta(n_v,m_v,\sd_v) \le A \biggl(m_v^{\tfrac{2s_v}{5s_v-4}}n_v^{\tfrac{5s_v-6}{5s_v-4}+\eps'}  + m_v^{2/3}n_v^{2/3+\eps'} + (m_v+n_v)\log m \biggr ),
\end{equation}
where $\eps'>\eps$ is an arbitrarily small constant $>\eps$, and $A$ is a sufficiently large constant that depends on $\eps'$ and on the other constant parameters.

The bound holds trivially for $\sd_v=1$ (the first term is `vacuous' for $\sd_v=1$). It also holds trivially for the case
$\sd_v>1$ and $n_v \le n_0$, if $A$ is chosen sufficiently large. The bound also holds when
$\sd_v>1$ and $m_v \ge n_v^{5/4}$ in view of Lemma~\ref{lem:primal}, again with a suitable choice of parameters.

The general case $\sd_v>1$, $m_v < n_v^{5/4}$, and $n_v > n_0$ is handled by double induction on $n$ and $\sd_v$. 
We omit the straightforward albeit somewhat tedious calculations; see~\cite{SZ} for a similar analysis (where an incidence bound was shown). 
Since $\sd_v \le s$, the total size of the biclique partition constructed by the algorithm, going back to the $O^*(\cdot)$ notation, is 
$O^*(m^{\tfrac{2s}{5s-4}}n^{\tfrac{5s-6}{5s-4}} + m^{2/3}n^{2/3}+m + n)$. A similar analysis shows 
that the expected running time of the algorithm is bounded by the same quantity (again, within the $O^*(\cdot)$ 
notation). This completes the proof of Theorem~\ref{thm:semioff} when each range is defined by one polynomial inequality.

\subsection{Handling general semi-algebraic ranges}
\label{app:general}

So far we have assumed that the regions in $\rangeS$ are defined by a single polynomial inequality. We next 
consider the case when they are defined by a conjunction of polynomial inequalities. That is, we assume that
we have a Boolean predicate $\Phi:\reals^2\times\reals^s \rightarrow \{0,1\}$ of the form
\begin{equation}
	\label{eq:conj}
	\Phi(\xx,\yy) = \bigwedge_{i=1}^k (g_i(\xx,\yy) \ge 0) ,
\end{equation}
where each $g_i\in\reals[\xx,\yy]$ is an $(s+2)$-variate polynomial of constant degree. Each 
$\srange_j\in\rangeS$ is  of the form $\srange_j = \{ \xx\in\reals^2 \mid \Phi(\xx,y_j) = 1\}$ for some 
$\yy_j\in\reals^s$. As above, we denote $\yy_j$ as $\tilde\srange_j$, and set 
$\tilde\rangeS = \{ \tilde\srange \mid \srange\in\rangeS\}$.  It will be convenient to think of computing 
a biclique partition of $\tilde\rangeS\mathop{\Phi} P$.

We compute such a biclique partition $\biclique$ by extending the idea in Section~\ref{sec:dual}. Namely, for $i\in [1,k]$,
let $\Phi_i:\reals^2\times\reals^s \rightarrow \{0,1\}$ be the Boolean predicate
\[
\Phi_i (\xx,\yy) = \bigwedge_{j=i}^k \left( g_j (\xx, \yy) \ge 0\right),
\]
and let $\phi_i (\xx,\yy)$ be the predicate $g_i(\xx,\yy)\ge 0$.
We compute a biclique partition $\biclique_i$ of $\tilde\rangeS \mathop{\Phi_i} P$ by recursing on $i$.  
Suppose we have computed 
$\biclique_{i+1}$; initially $i=k$ and we set, vacuously,  $\biclique_{k+1} = \{(\rangeS,P)\}$.
Let $(\tilde\rangeS_j, P_j) \in \biclique_{i+1}$.
We compute a biclique partition $\biclique_{ij}$ of $\tilde\rangeS_j \mathop{\phi_i} P_j$,
using the algorithm of Section~\ref{subsec:algo}, and set $\biclique_i= \bigcup_j \biclique_{ij}$.

Each recursive subproblem is now defined by a 4-tuple $(\F_v, \tilde\rangeS_v, P_v, i)$, where 
$\tilde\rangeS_v \subset Z(\F_v)$, and the goal is to compute a biclique partition $\biclique_v$ 
of $\tilde\rangeS_v \mathop{\Phi_i} P_v$. 
We follow the same approach as above, but there are now three threads of recursion. Two of the threads
are the same as above. 
For each cell $\cell\in \Xi$, let $(\tilde\rangeS_\cell, P_\cell^\circ)$ be the biclique as defined above. 
If $i=k$ then we add $(\tilde\rangeS_\cell, P_\cell^\circ)$ to $\biclique_v$. Otherwise ($i<k$), we 
recursively solve the problem $(\F_v, \tilde\rangeS_\cell, P_\cell^\circ, i+1)$. 
We obtain a similar recurrence as above. In particular, for the general case $n>n_0$, $i<k$, and $\sd>1$, 
we obtain the following recurrence:
\begin{align*} 
\beta(n_v, m_v, s_v, i) \le 
	& \sum_{\cell\in \Xi} \beta(n_\cell,m_\cell,s_v,i) + \sum_{\cell\in \Xi} \beta(n_\cell,m_v,s_v,i+1)+
		\beta(n_{v,0},m_v,s_v-1,i) .
\end{align*}
The solution of this recurrence, using an additional induction on $i$, is also 
$$O^*\biggl (m^{\tfrac{2s}{5s-4}}n^{\tfrac{5s-6}{5s-4}} + m^{2/3}n^{2/3}+m + n\biggr),$$
as is easily verified.

Following a standard approach, as outlined in~\cite[Appendix A]{AAEKS22},
we note that our algorithm can be extended in a straightforward manner 
to compute a biclique partition for the predicate $\neg\Phi(\xx,\yy)$, where $\Phi$ is a predicate of the form (\ref{eq:conj}). 
Finally, suppose $\Phi$ contains a disjunction, i.e.,
$\Phi(\xx,\yy) = \Phi_1(\xx,\yy)\vee \Phi_2(\xx,\yy)$. Then we first compute a biclique partition
$\biclique_1$ of $\tilde\rangeS \mathop{\Phi_1} P$, and then compute a biclique partition $\biclique_2$ of 
$\tilde\rangeS \mathop{(\neg\Phi_1\wedge\Phi_2)} P$, again using the machinery outlined in~\cite[Appendix A]{AAEKS22}. We return $\biclique_1 \cup \biclique_2$ as the desired biclique partition.
This completes the proof of Theorem~\ref{thm:semioff}.

Finally, we remark that if each polynomial inequality $g_i(\xx,\yy)\ge 0$ in the definition of $\Phi$ uses at 
most $\bar{s}$ variables of $\yy$, for some $\bar{s}\le s$, then the hierarchical partition in 
Section~\ref{subsec:algo} is constructed \footnote{More precisely, each level in the hierarchy can be implemented in $\reals^{\bar{s}}$, although these subspaces capture different subsets of the $s$ parameters specifying $\yy$.}
in $\reals^{\bar{s}}$ instead of $\reals^s$, and the size of the biclique partition becomes
$$O^*\biggl (m^{\tfrac{2\bar{s}}{5\bar{s}-4}}n^{\tfrac{5\bar{s}-6}{5\bar{s}-4}} + m^{2/3}n^{2/3}+m + n\biggr),$$
which can be much smaller in some cases. For example, if $\Sigma$ is a set of triangles in $\reals^2$, then $s=6$, 
while standard simplex range searching machinery uses only $\bar{s}=2$. See~\cite{AAEKS22} for further details.

\section{Conclusion}

In this paper we presented efficient algorithms for answering semi-algebraic range queries and 
point-enclosure queries in the plane in an off-line setting. In particular,
given a set $P$ of $m$ points in $\reals^2$ and a set $\rangeS$ of $n$
semi-algebraic sets of constant complexity in $\reals^2$, we presented a randomized algorithm for 
computing a biclique partition $\biclique$
of size 
\[
O^*\bigl( m^{\frac{2s}{5s-4}}n^{\frac{5s-6}{5s-4}} + m^{2/3}n^{2/3} + m + n \bigr)
\]
of $\rangeS\mathop{\Phi} P$, where $s>0$ is the number of degrees of freedom of the regions in $\rangeS$. 
It is straightforward to answer both range and point-enclosure queries, 
in either off-line or on-line manner,
using $\biclique$ (in the online setting, the queries come only from the prescribed set $\Sigma$ or $P$).

A recent result of Chan~\etal~\cite{CCZ24} shows that $m$ point-enclosure queries amid a set 
of $n$ semi-algebraic sets in $\reals^2$, in an on-line setting, can also be answered in 
\[
O^*\bigl( m^{\frac{2s}{5s-4}}n^{\frac{5s-6}{5s-4}} + m^{2/3}n^{2/3} + m + n \bigr)
\]
expected time. Hence, the time complexity of answering two-dimensional point-enclosure queries is the same 
(within a subpolynomial factor) in both off-line and on-line settings. The approach in \cite{CCZ24}, however, 
does not extend to on-line semi-algebraic range queries in $\reals^2$, and thus there is a gap between off-line and 
on-line semi-algebraic range searching in $\reals^2$.
The most natural (and apparently deep) open question is to bridge this gap.

Another interesting question is whether our technique can be 
extended to off-line semi-algebraic range queries in $\reals^3$. In particular, let $P$ be a set of $m$ 
points in $\reals^3$ and $\rangeS$ a set of $n$ semi-algebraic sets in $\reals^3$ with $s$ degrees of freedom.
Using standard techniques, reviewed in the Related work part of the introduction, one can construct a biclique partition of $\rangeS\mathop{\Phi} P$ 
of size $O^*\bigl(m^{\tfrac{2s}{3s-1}}n^{\tfrac{3s-3}{3s-1}}+m+n \bigr)$. 
Can this bound be improved in an off-line setting?

\paragraph{Acknowledgements.}
We thank Nabil Mustafa and Sergio Cabello for useful discussions that motivated the study
reported in this paper.

\end{document}